%% file: main.tex
\documentclass[runningheads]{llncs}

\newcounter{noteDMctr} 

\newif\ifEditMode

\EditModetrue

\usepackage{aliases}
\usepackage{preamble}

\begin{document}
\title{Quantifying Arbitrage in \\Automated Market Makers: \\An Empirical Study of Ethereum ZK Rollups}
%
\titlerunning{Quantifying Arbitrage in AMMs}
%
\author{Krzysztof Gogol\inst{1,5} \and
Johnnatan Messias\inst{5} \and
Deborah Miori \inst{3} \and
Claudio Tessone  \inst{1,4}  \and
Benjamin Livshits \inst{2,5}}

\authorrunning{K. Gogol, J. Messias, D. Miori, et al.}

\institute{University of Zurich \and
University of Oxford \and
Imperial College London \and
UZH Blockchain Center \and
Matter Labs }



\maketitle
\begin{abstract}
    \input{sections/00Abstract.tex} \\
    \noindent\textbf{Keywords:} Arbitrage, Automated Market Maker, Rollup
\end{abstract}
    

\input{sections/01Introduction}
\input{sections/02Background}
\input{sections/03Theory}
\input{sections/04Empirical}
\input{sections/05RegressionModel}
\input{sections/09Conclusions}

\section*{Acknowledgements}  
This research article is a work of scholarship and reflects the authors' own views and opinions. It does not necessarily reflect the views or opinions of any other person or organization, including the authors' employer. Readers should not rely on this article for making strategic or commercial decisions, and the authors are not responsible for any losses that may result from such use. 
One of the authors' research was supported by the \textit{EPSRC CDT in Mathematics of Random Systems} (EPSRC Grant  EP/S023925/1).

\bibliographystyle{splncs04}
\bibliography{main}
	
\appendix
\input{sections/97AppendixTheory}

\end{document}

%% file: sections/00Abstract.tex


Arbitrage opportunities arise from the simultaneous buying and selling of the same asset in different markets to profit from price differences. This research systematically explores arbitrage possibilities between Automated Market Makers (AMMs) on Ethereum zk-rollups and Centralized Exchanges (CEXs). To start, we introduce a theoretical framework to assess arbitrage opportunities and develop a formula for the Maximal Arbitrage Value (MAV), considering both price discrepancies and the liquidity present in trading venues. 

Following this, we conduct an empirical assessment of the historical MAV between SyncSwap, an AMM on zkSync Era, and Binance, examining the speed at which price discrepancies are resolved considering both explicit and implicit market costs. Overall, the total MAV from July to September~2023 in the USDC-ETH SyncSwap pool amounts to~\$104.96K~(0.24\% of trading volume). 

%% file: sections/01Introduction.tex

\section{Introduction}
    \label{sec:intro}

Decentralised Finance (DeFi) presents a novel alternative to traditional financial services \cite{werner2022sok, schaer2023defimarkets, Auer2023technologydefi, miori2024clustering, miori2023defi}. Unlike conventional finance, DeFi operates without intermediaries, providing financial services directly through the blockchain network. However, the promise of democratisation inherent in DeFi has been hindered by the high gas costs associated with blockchain transactions. 

The Blockchain Trilemma, formulated by V. Buterin, the founder of Ethereum, states that each blockchain can prioritise only two out of three factors: decentralisation, scalability, or security \cite{werth2023review}. Given that decentralisation is fundamental for blockchain, and security is paramount for most applications, Ethereum prioritises these two elements. Consequently, this first programmable blockchain experienced network congestion and a surge in gas fees, hampering its adoption and the use of DeFi, which predominantly operates there. Indeed, the Ethereum ecosystem accounts for 51\% of the trading volume and 63.24\% of Total Value Locked (TVL) in DeFi \cite{2024DeFiLlama}.
To tackle this challenge, various scaling solutions have been proposed. Layer-1 (L1) scaling involves the development of a new blockchain with its own consensus mechanism and infrastructure. 
Then, Layer-2 scaling (L2), among which most common are rollups, relies on the security and decentralisation of the underlying network, typically Ethereum, but offloads complex computations outside of it to store there only results \cite{sguanci2021layer, gangwal2022survey}. 
Recently, we have observed a shift in TVL and trading volumes from Ethereum to its rollups. Arbitrum, Optimism (optimistic rollups), and zkSync Era (zk-rollup) tended to process indeed more transactions compared to Ethereum in 2023. At the time of writing, Arbitrum ranks fourth and zkSync Era ranks seventh in terms of trading volume, surpassing older L1 blockchains \cite{2024DeFiLlama}. This shift has been encouraged by the fact that, on average, rollups offer 50 times lower gas fees compared to the underlying chain \cite{yee2022shades}.

In an efficient market, token prices at Centralised Exchanges (CEXs) and Decentralised Exchanges (DEXs) at any L1 and L2 blockchains should be equal. However, misalignment of prices is observed to commonly persist \cite{barbon2023quality} and can thus lead to arbitrage opportunities. 
In this study, we evaluate the efficiency of DEXs on zkSync Era, which is the largest venue in terms of trading volume and capital among Ethereum zk-rollups. We focus on rollups due to their enhanced speed of block production (two seconds) compared to Ethereum mainnet (twelve seconds) and lower fees.
In particular, we investigate the convergence time between prices of DEXs on zkSync Era and Binance, assessing the related arbitrage opportunities also with respect to the liquidity available in the systems.


This study also contributes to the research on rollup design, and particularly on the current discussions about the enablement of certain forms of Maximal Extractable Value (MEV) by miners or validators on L2s \cite{mamageishvili2023BuyingTime, mamageishvili2023shared}.
\subsubsection*{Contributions.} Our work analyzes arbitrage opportunities that arise from price misalignments between layer-2 AMMs and crypto CEXs. 

\begin{itemize}
\item We introduce a framework to record and assess these arbitrage cases, and to determine their decay period (i.e., the duration until price re-alignment). This system modifies the net LVR metric~\cite{milionis2023automated} using an empirical approach to calculate arbitrage profit when price differences continue across multiple blocks, thus avoiding the repeated counting of identical opportunities.

\item We present \emph{Maximal Arbitrage Value} (MAV) as a metric that captures the total profit obtainable from a specific price discrepancy and the existing liquidity in the trading platforms. Subsequently, we conducted a test scenario and computed MAV at zkSync Era for the highly active USDC-ETH trading pool and observed that the arbitrage opportunities lasted for several minutes before diminishing.

\item We break down and estimate the costs (i.e., AMM fees, gas fees, and block slippage) that arbitrageurs would need to sustain to execute the desired swap transaction. This enables us to derive new understandings of the dynamics that define arbitrage opportunities between DEXs and CEXs, as well as implications for market design. We observed negligible block slippage, including MEV, affecting swaps in the examined pool.
\end{itemize}

\subsubsection*{Paper Organization.} 
Section \ref{sec:background} provides the necessary background knowledge on rollups, DEXs and CEXs.
Section \ref{sec:theory} proposes our theoretical framework for investigating arbitrage between DEX and CEX, while Section \ref{sec:evidence} describes an empirical case study of such arbitrage opportunities. 
Section \ref{sec:model} dives into the arbitrage instances we identified and models their fluctuation in time based trading preferences and market fees.
Section~\ref{sec:discussion} provides a discussion of the role of arbitrage. 
Finally, Section \ref{sec:conclusions} concludes our work with some final remarks and directions for future research.


%% file: sections/02Background.tex
\section{Background: Rollups, DEXs and CEXs}
\label{sec:background}

In this section, we first introduce the concept of L2 blockchain scaling (especially rollups) and then proceed to provide useful background information on both crypto CEXs and DEXs.

\subsection{Layer-2 and Rollups}

There are two approaches that address the blockchain scalability challenges and the related spikes in gas prices: Layer-1 (L1) and Layer-2 (L2) scaling. The L1 scaling involves the creation of an entirely new blockchain, with new consensus mechanisms and distinct physical infrastructure that maintains the network.
L2 scaling adopts an alternative strategy: computations are executed outside of the main blockchain, but their results are saved in the underlying chain. The major types of L2 include state (payment) channels, plasma, and rollups. While plasma and state channels represent L2s aiming to move both data and computation off-chain, rollups are non-custodial: they move computation and state storage off-chain but retain compressed data per transaction on the underlying chain. 

\begin{figure}[!htb]
  \centering
        \includegraphics[width=0.6\linewidth]{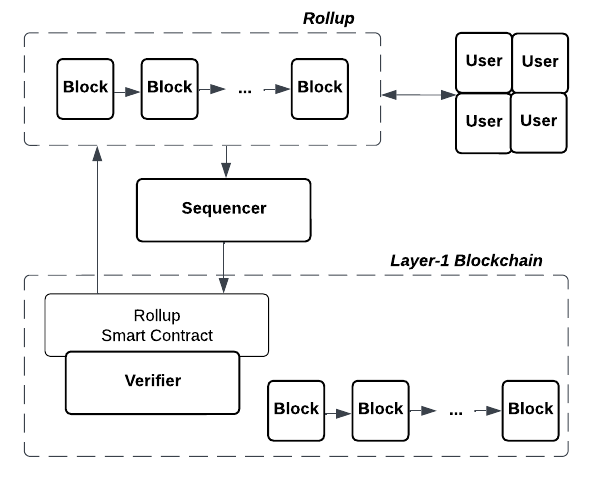} 
    \caption{Architecture of a rollup.}
    \label{fig:rollup-diagram}
\end{figure}

A rollup functions as a blockchain: it produces blocks and snapshots those blocks to the main chain. However, it operates in an environment where operators are not trusted, implying that operators can potentially act adversarially by stopping block production, producing invalid blocks, withholding data, or attempting other malicious behaviors. There are two approaches to ensure the correctness of the state of rollups, denoted as optimistic and zero-knowledge proofs~\cite{sguanci2021layer, gangwal2022survey}.

Figure~\ref{fig:rollup-diagram} illustrates the rollup architecture with its key components, which are sequencers and verifiers. Sequencers roll up transactions to the L1 chain and, by bundling such transactions, they generate great savings on gas fees. Verifiers are smart contracts that operate on L1 and verify the transactions stored by the sequencer, and they ensure the correctness of the transactions~\cite{sguanci2021layer, yee2022shades}.
Unlike in L1 blockchains, gas fees are not known upfront for transactions in rollups. This comes from the fact that there are two elements within such fees, namely, gas fees charged by the sequencer and gas fees charged by the L1 network. First, the gas fee for a transaction is estimated by the sequencer, and the address that initiated the transaction is charged the estimated gas cost. Once the transaction (batched with other rollup transactions) is stored within the layer-1 chain, the final gas fees from L1 are known. Subsequently, the gas fee overpaid by the initial transaction originator is refunded.

\subsection{Crypto Centralised Exchanges}


Crypto Centralised Exchanges (crypto CEXs) allow to convert fiat currencies, e.g. USD or EUR, into cryptocurrencies, e.g. ETH, BTC, USDC, and back. Traders often have the option to decide whether they want to custody their cryptocurrencies at the crypto CEX, or transfer them to their wallets at the selected L1 or L2 blockchains. 
Crypto CEXs operate similarly to traditional stock exchanges and employ Limit Oder Books~(LOBs) that store all open buy and sell transactions (limit orders) and match the incoming market orders. Designated market makers, which typically are financial institutions, provide capital to LOBs to facilitate smooth trading.

\subsection{Decentralised Exchanges}


DEXs allow traders to swap tokens directly on the blockchain in a non-custodial and atomic way. They usually leverage the AMM framework to automatically determine exchange rates between tokens from the relative reserves of the traded tokens within the liquidity pool. This framework (i.e. AMMs) was first introduced as a solution to the prohibitive storage costs that prevent the implementation of a LOB set-up on blockchains, and is becoming increasingly accepted by the financial communities.
\paragraph{Automated Market Makers.}
Constant Function Market Makers are the dominant type of AMMs. They are equipped with a reserve curve function $F: \mathbb{R}^N_+ \rightarrow \mathbb{R}$ that maps $N$ token reserves $x_i$ to a fixed invariant $L$. The latter is determined by the amount of tokens locked in the pool and provided by Liquidity Providers~(LPs). The reserve curve functions are designed to maximize capital efficiency within the liquidity pool~\cite{Xu_2023}.
Each swap (i.e. exchange of two tokens) at an AMM incurs a trading fee (also called LP fee), usually in the form of a fixed percentage of the swap that must be paid by the trader. These fees, or a portion thereof, are then distributed proportionally among the LPs in the pool to reward them for the service provided by locking their tokens into the pool.

\paragraph{Constant Product Market Makers (CPMM).} Uniswap v2 is one of the first and most popular AMMs that are CPMMs \cite{Adams2020UniswapCore}. Its trading is regulated by the formula
\begin{equation}\label{eq:g_uniswapv2}
x_1 \cdot x_2 = L, 
\end{equation}
where $x_i$s are the balances of the two tokens characteristic of each liquidity pool, and $L$ is a related constant signalling the level of liquidity of the pool.

\paragraph{CPMM with Concentrated Liquidity (CLMM).} Concentrated liquidity is an enhancement to CPMMs introduced within Uniswap v3 \cite{Adams2021UniswapCore} to increase the efficiency of liquidity locked in a pool. It allows LPs to specify a price range on which liquidity is provided. The trade invariant for the two-token pool $N=2$ in the price interval $[p_j,p_k]$ is then
\begin{equation}\label{eq:g_uniswapv3}
\left( x_1+\frac{L}{\sqrt{p_k}} \right)\left(x_2+L\cdot \sqrt{p_j}\right) = L^2,
\end{equation}
where variables have similar meanings as before.
The disadvantage of CLMMs is that LPs must constantly monitor and update the price range on which they provide liquidity, since only swaps occurring within such a range make them profit from LP fees. Thus, LPs need to devise a strategy to avoid incurring undesirable levels of gas fees. The \emph{Stableswap Invariant}, introduced with Curve v1~\cite{Egorov2019StableSwapLiquidity}, automatically concentrates liquidity around the current price. However, it is specifically designed for tokens that trade at a constant price with respect to each other, such as two stablecoins pegged to the US dollar. \emph{Cryptoswap Invariant}, introduced with Curve v2~\cite{Egorov2021CurvePeg}, is then an adaptation of the Stableswap Invariant for any token pairs.



%% file: sections/03Theory.tex
\section{The Theory of Arbitrage on AMMs}
\label{sec:theory}

Price disparities between multiple venues (either DEXs and/or CEXs) can lead to interesting arbitrage opportunities. In this section, we derive the transaction volume that fully re-aligns prices, given the current token reserves in an AMM liquidity pool. By trading such volume, an arbitrageur's profit is consequently maximized, whose value we refer to as Maximal Arbitrage Value~(MAV). Importantly, due to much higher reserves usually available in crypto CEXs versus DEXs, we assume for simplicity that trading on the CEX has no price impact.

\subsection{Maximal Arbitrage Value}


Let us assume that a price difference exists between a CEX and an AMM, and further assume that the CEX has much higher market depth than the AMM (i.e. we have zero market impact on its side).
Let $P_c$ and $P_a$ be the prices on the CEX and AMM, respectively, and assume that $P_c$ < $P_a$.
We are looking for the maximum trade volume $V_{max}$ that we can buy at the CEX for $P_c$ and sell at the AMM for $P_a$.
So, $V_{max}$ is the trade volume that equals the prices at AMM and CEX, given \emph{percentage price impact} $\rho (V_{max})$ imposed upon trading. Thus, we can define our MAV as


\begin{equation*}
MAV = V_{max} \cdot \Big( P_a \cdot ( 1 - \rho( V_{max})) -  P_c \Big),
\end{equation*}
or equivalently
\begin{equation}
MAV = V_{max} \cdot ( P_a  -  P_c ) - V_{max} \cdot P_a \cdot \rho( V_{max}).
\end{equation}

\paragraph{Arbitrage between a CPMM and CEX.} Clearly, we now need to derive what the percentage price impact $\rho(V_{max})$ of a given volume $V_{max}$ is on the AMM. 
Let us assume that we have two tokens $X$ and $Y$, with the respective reserves in the liquidity pool of AMM being $x_1=x$ and $x_2=y$.
Assuming we are dealing with a CPMM, then the act of trading follows the $x_1 \cdot x_2 = x \cdot y = L$ formula, and the instantaneous exchange rate $P_a$ for an infinitesimal trade is
\begin{equation}
    P_a = \lim_{\Delta y \rightarrow 0}  \frac{\Delta x}{\Delta y} = - \frac{\partial}{\partial y} \Big ( \frac{L}{y} \Big ) = \frac{L}{y^2} = \frac{x}{y}.
    \label{exchange-rate}
\end{equation}
The \textit{execution rate $\Tilde{P_a}(\Delta y)$} that a trader obtains when selling a quantity $\Delta y > 0$ is then
\begin{equation}
    \Tilde{P_a}(\Delta y) = - \frac{\frac{L}{y+\Delta y} - \frac{L}{y}}{\Delta y},
\label{exec-rate}
\end{equation}
and so the percentage price impact $\rho(V_{max} = \Delta y)$ can be defines as
\begin{equation}
    \rho = \frac{\Tilde{P_a}(\Delta y)}{P_a} - 1.
\label{percentage price impact}
\end{equation}
Then, we re-write
\begin{align*}
    \rho(\Delta y) &= \frac{y}{x} \cdot \Big( \frac{-\frac{xy}{y+\Delta y} + \frac{xy}{y}}{\Delta y} - \frac{x}{y} \Big) \\
    &= \frac{1}{x} \cdot \Big( \frac{\frac{-xy^2}{y+\Delta y} + xy - x \Delta y}{\Delta y} \Big) = \frac{1}{x} \cdot \Big( \frac{\frac{+xy\Delta y}{y+\Delta y} - x \Delta y}{\Delta y} \Big) \\
    &= \frac{1}{x} \cdot \Big(\frac{+xy}{y+\Delta y} - x \Big) = \frac{1}{x} \cdot \Big(\frac{- x\Delta y}{y+\Delta y} \Big).
\end{align*}

By remembering that $(x+\Delta x)\cdot (y + \Delta y) = xy$, we can further write $x\Delta y + y\Delta x + \Delta x \Delta y = 0$. Thus, $x\Delta y = -\Delta x (y+ \Delta y)$ and
\begin{equation}
    \rho(V_{max} = \Delta y) = \frac{\Delta x}{x}.
\end{equation}
Thus, we have just shown that the percentage price impact on a CPMM (such as Uniswap v2) is given by $\Delta x / x $ for a swap of $\Delta x $ for $\Delta y $.
Consequently, our MAV definition can be further simplified to 
\begin{equation}
\label{eq:MAV_general_uniswap_v2}
MAV = V_{max} \cdot ( P_a  -  P_c ) - \frac{V_{max}^2 P_a}{x},
\end{equation}
but we are still in need for a computation of the $V_{max}$ optimal to be traded. Thus, we compute the first derivative of Eq. \eqref{eq:MAV_general_uniswap_v2} with respect to $V_{max}$ and force it to zero to find any turning points. This is
\begin{equation*}
0 = ( P_a  -  P_c ) - 2 \frac{V_{max} P_a}{x}
\end{equation*}
and implies that
\begin{equation}
\label{eq:V_max_uniswap_v2}
V_{max} =  \frac{y \cdot ( P_a  -  P_c )}{2 P_a}.
\end{equation}
Finally, by substituting \eqref{eq:V_max_uniswap_v2} into \eqref{eq:MAV_general_uniswap_v2} we can thus fully characterise our MAV formulation by writing
\begin{equation}
\label{eq:MAV_uniswap_v2}
MAV = \frac{y \cdot ( P_a  -  P_c ) ^ 2 }{4 P_a}.
\end{equation}

%% file: sections/04Empirical.tex
\section{Empirical Case Study: zkSync Era}
\label{sec:evidence}

Whenever there is a price divergence between trading venues, arbitrageurs can swiftly capitalize on the opportunity and try to profit from re-aligning prices. This dynamic benefits both arbitrageurs, who indeed profit from exploiting the price difference, and conventional traders, who benefit from quickly updated and accurate exchange rates. 

In this section, we analyze the price differences between AMM pools on zkSync Era versus Binance (currently the most important crypto CEX). We measure the decay time of these disparities and quantify emerging arbitrage opportunities using our proposed MAV metric. Subsequently, we study the impact of trading fees, block slippage, and gas costs on the arbitrage profit and decay time of such price misalignments.

\subsection{Dataset We Used}
\label{sec:dataset}

We obtain the transaction log data from the archive node of zkSync Era covering all swaps made at major DEXs at zkSync Era (i.e. zkSwap and syncSwap) during the period from July 1st, 2023 to September 17th, 2023.
We find that the largest pools, in terms of TVL and trading volume, are:
\begin{itemize}
    \item SyncSwap USDC-WETH (5,933,132 swap transactions),
    \item SyncSwap WETH-WBTC (144,834 swap transactions),
    \item zkSwap USDC-WETH (510,270 swap transactions),
    \item zkSwap WETH-WBTC (5,869 swap transactions),
\end{itemize}
in all of which a trader needs to pay $8$bps of fees of any volume desired to be exchanged.

For each swap transaction, the data set includes transaction logs from both the \emph{Swap method}, i.e. token amounts in and out, and from the \emph{Sync method}, i.e. token balances in the liquidity pool. From the Sync method, we retrieve the liquidity pool balance before the swap occurs. Based on \texttt{transactionIndex} and \texttt{logIndex}, we identify the order of swap transactions within the block on zkSync Era in order to identify the block slippage and possible adversarial change in swap rates. The CEX exchange rates between USDC-ETH and ETH-BTC are sourced from Binance; we use one-minute time intervals.

Throughout all our following analyses, we focus on the most active pool (i.e. the SyncSwap USDC-WETH pool), since it clearly is the most representative of a healthier trading environment.

\subsection{Proposed Analysis Framework}

\begin{figure}[h!]
\begin{subfigure}{.69\textwidth}
  \centering
    \includegraphics[width=0.95\linewidth]{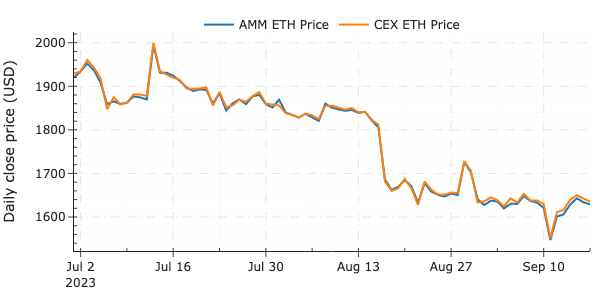}
    \caption{Time series of prices for USDC-WETH.}
    \label{fig:price-diff-evidence}
  \end{subfigure}
  \begin{subfigure}{.3\textwidth}
  \centering
    \includegraphics[width=0.95\linewidth]{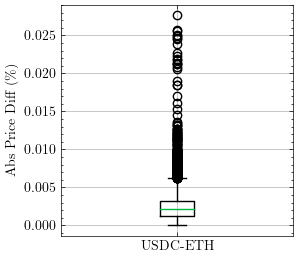}
    \caption{Distribution of deltas.}
    \label{fig:delta-price-outliers}
  \end{subfigure} \\
  \vspace{0.15cm} \\
  \centering
  \begin{subfigure}{.8\textwidth}
  \centering
    \includegraphics[width=0.99\linewidth]{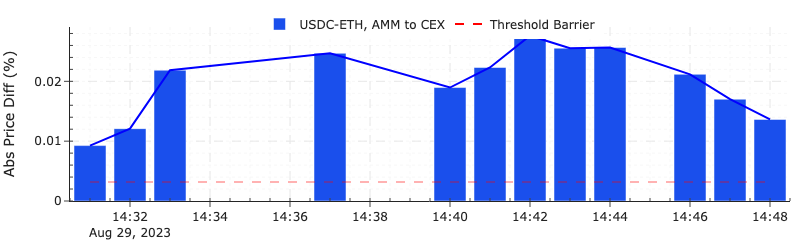}
    \caption{Price differences at minute level. The red dashed line represents the threshold at which we assume that prices have re-aligned. Missing bars indicate that no trading has happened on the AMM that minute.}
    \label{fig:explanatory-decay}
  \end{subfigure}
  \caption{Explanatory plots for the price differences for USDC-ETH among SyncSwap AMM on zkSync Era and Binance. The AMM spot price oscillates around the CEX price and we can notice how misalignments expand and then decay over several minutes.}
\end{figure}

We now propose a framework for empirically identifying and investigating the arbitrage opportunities mentioned above. This framework is composed of the following elements:
\begin{enumerate}
    \item \textbf{Time Intervals}: For each minute, we compare the CEX closing price with the latest spot price on DEX. The spot price at DEX is calculated based on the token balances in the liquidity pool after the last swap in the minute. Figure~\ref{fig:price-diff-evidence} presents the ETH-BTC exchange rates at Binance and SyncSwap, and it can be observed that there are price differences.
    
    \item \textbf{Threshold for Price Re-Alignment}: We consider only price differences that are larger than a pre-defined threshold. This threshold is chosen from the empirical distribution of price disparities (see Figure~\ref{fig:delta-price-outliers}), and we primarily focus on outlier quantities (i.e. outside~1.5 times the interquartile range above the third quartile).
    In particular, the chosen threshold parameter implies that we consider prices to be aligned for the majority of time, but we can modify it to allow for more flexibility.
    
    \item \textbf{Empirical MAV}: Figure~\ref{fig:explanatory-decay} shows the magnitude of the price misalignments. As the price difference persists over a long period of time, we save only the maximum MAV within each time interval with price misalignment until the next re-alignment. With this approach, we avoid double-counting MAV for the same evolving price misalignment.
    \item \textbf{Decay Time}: After identifying all occurrences of empirical MAV (i.e. maximum MAV during a price misalignment slot), we calculate the time until prices re-align (i.e. when the price difference in magnitude falls below our chosen threshold).
\end{enumerate}

\subsection{Empirical MAV from the USDC-ETH Pool}

\begin{figure}[h!]
  \centering
        \includegraphics[width=0.9\linewidth]{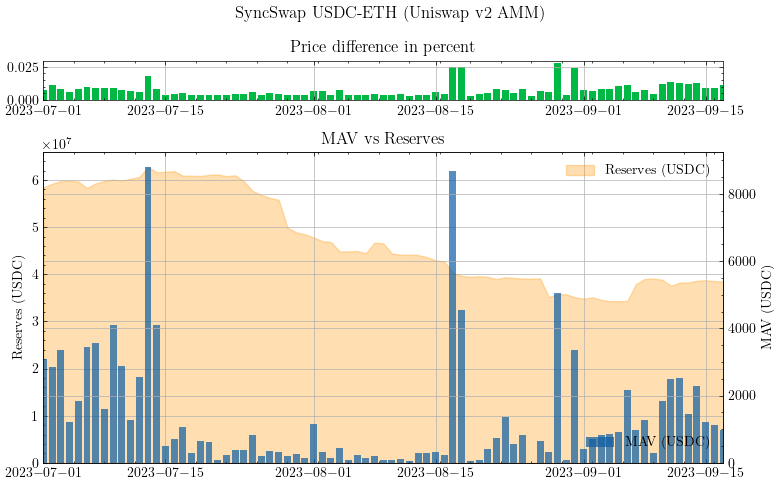} 
    \caption{Empirical MAV, price misalignment and reserves in the USDC-ETH pool at SyncSwap on zkSync Era.}
    \label{fig:MAV_SyncSwapUSDCETH}
\end{figure}

As just highlighted, a bigger opportunity for arbitrage profit arises if there is a stronger divergence of prices. Still, it also depends on the market impact of the trade (i.e. reserves of the pool). Figure~\ref{fig:MAV_SyncSwapUSDCETH} presents price differences over time, empirical MAV, and reserves. In particular, it shows the maximum (in absolute value) price difference over each day, the daily sum of empirical MAV found that day, and the last TVL value on that day (i.e. the sum of values of the two tokens' reserves on a common basis).

The daily MAV ranges from~$0 to~$8,000, with the minimum values occurring during a phase of low trading volume from mid-July to early September~2023. Additionally, the reduction in token reserves within the pool leads to a lower MAV throughout the analyzed period. Moreover, the days with the highest daily MAV (13/7, 14/7, 17/8, 18/8, 29/8, 31/8) correspond to the days with the largest price disparities. Noticeably, even though the largest price disparity is observed on August~18th, the highest daily MAV is observed on July~13th, due to the higher TVL in the liquidity pool reserves on this day (i.e. more volume can be traded to obtain the same market impact).

Throughout the analyzed timeframe, a trading volume of~\$43.73M resulted in a cumulative MAV of~\$104.96K, equating to~0.2349\% of the total trading volume.

\color{black}


\subsection{Trading Fees}
\label{sec:fees}

The total expense of performing swap transactions on an AMM includes gas fees as well as explicit and implicit DEX charges~\cite{Xu_2023, gudgeon2020defi}. For clarity, we write

\begin{align*}
\label{eq:cost}
Total Fees = \underset{Gas Fee}{\underbrace{L1 Fee + L2 Fee}} + LP Fee + Block Slippage,
\end{align*}
and now dive into them.
\\
\textbf{Gas Fees in Rollups.} In contrast to L1 blockchains, gas fees for rollup transactions are not predetermined, as they consist of two parts: gas fees imposed by the rollup sequencer (\emph{L2 Fees}) and gas fees imposed by the L1 network (\emph{L1 Fees}). Initially, the sequencer estimates the gas fee for a transaction, and the address initiating the transaction is charged this estimated amount. When the transaction, along with a batch of other rollup transactions, is recorded on the L1 chain, the actual L1 gas fees become known. Depending on the rollup implementation, any overpaid gas fees are either refunded to the transaction originator (e.g., zkSync Era, which is examined in the research) or deducted from future transaction costs (e.g., Arbitrum).
\\
\textbf{LP Fee.} They are also called trading fees, and represent the explicit fee paid by the trader to DEX to swap the tokens. They typically range between~1 and~50 bps of the trade volume and are redistributed to LPs.
\\
\textbf{Block Slippage.} This represents a hidden cost of swapping at an AMM, potentially resulting in an unanticipated execution price due to the sequence in which transactions are processed within a block.~\cite{werner2022sok,schaer2023defimarkets,Auer2023technologydefi,Gogol2023SoK:Risks}. 
The impact of slippage can be either positive or negative, depending on whether its impact on the executed price is favorable or unfavorable to a trader. Currently, transaction re-ordering~(MEV) attacks are not possible at rollups because rollups, including zkSync Era, presently operate a centralized sequencer. Additionally, zkSync Era generates new transaction blocks every~2 seconds, whereas Ethereum produces them every~12 seconds, thereby reducing the likelihood of block slippage. 


%% file: sections/05RegressionModel.tex
\section{Insights Into MAV Opportunities Identified}
\label{sec:model}

We consider all empirical MAV events for price divergences that correspond to outliers against the distribution of misalignments, and their time decay (i.e. see Figure~\ref{fig:MAV_SyncSwapUSDCETH}). 
We would expect that the higher the MAV is, the quicker the associated decay follows, but decide to test this intuition throughout.
For each identified instance of MAV to possibly profit from, we estimate the costs that an arbitrageur needs to consider before trading (i.e. fees to liquidity providers, gas fees, and block slippage due to congestion) as described in Section \ref{sec:fees}.
Indeed, the set of different market fees that an arbitrageur needs to pay does also intuitively impact the speed of decay of price misalignments and of exhaustion of MAV opportunities.

In this section, we thus complete two complementary analyses. We first cluster the identified empirical MAV events, in order to extract \say{ordinary} scenarios, while detecting an extremely unusual combination of features that we also discuss. Subsequently, we concentrate on typical MAV events and examine the statistical correlations that illustrate how the decay time depends on both MAV magnitude and fees.



\subsection{Clustering Analysis}

We found $116$ instances of MAV to profit from within our most active pool (i.e. trading USDC-ETH).
For each one of those instances, we considered multiple factors, such as block slippage, trading volume, number of swaps and cumulative gas until the decay, and found the following set of features significant:
\begin{enumerate}
    \item \texttt{time\_decay} - Witnessed length of time until the alignment of prices is restored (in seconds).
    \item \texttt{clean\_MAV} - Empirical MAV identified, but subtracting the deterministic LP fees that the arbitrageur needs to pay to trade. Nicely, all MAV events are still positive in profit after this discount and are thus further considered within our analyses.
    \item \texttt{avg\_gas} - Average gas fees per swap transaction witnessed by the possible arbitrageur within the one minute before he/she wants to trade to earn the available MAV.
    \item \texttt{Vmax\_on\_usage} - Ratio between the MAV volume $V_{max}$ that the arbitrageurs would like to trade and the overall volume traded on the venue during the preceding minute. This feature acts as a proxy for the risk of the arbitrage trade to be impacted by external \texttt{noise} (and possibly adversarial) trading.
\end{enumerate}
We then complete Principal Component Analysis on the standardized version of the above data, and find that the four eigenvalues account for~$\{32,26,24,18\}$ \% of the variance, respectively. Thus, there is no single main component against which the data could be directly summarized.

\begin{figure}[th]
  \begin{subfigure}{.5\textwidth}
  \centering
    \includegraphics[width=1\linewidth]{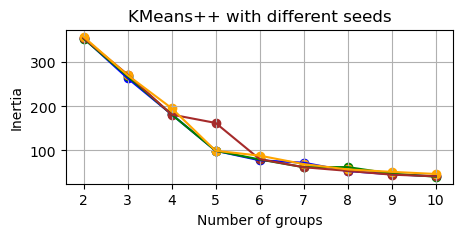}
    \caption{Inertia plot pointing to best clustering.}
    \label{fig:inertia-plot}
  \end{subfigure}%
  \begin{subfigure}{.5\textwidth}
  \centering
    \includegraphics[width=0.8\linewidth]{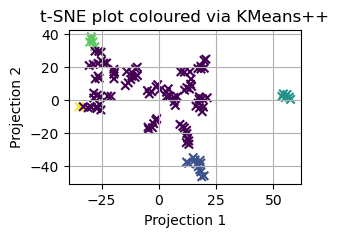}
    \caption{t-SNE projection of MAV events.}
    \label{fig:tsne}
  \end{subfigure} \\
  \centering
  \begin{subfigure}{.75\textwidth}
  \centering
    \includegraphics[width=1\linewidth]{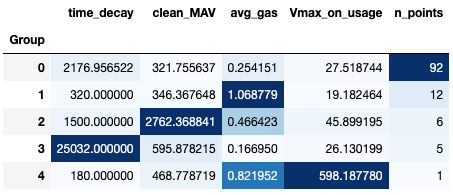}
    \caption{Average features for each cluster identified by KMeans++.}
    \label{fig:summary-stats}
  \end{subfigure}
  \caption{We cluster our data points relating to MAV events. The inertia plot suggests an optimal clustering into~$5$ groups, which are visualised via a t-SNE projection and analysed in the proposed table.}
\end{figure}

We proceed to cluster our data via KMeans++ and experiment with different seeds to initiate the algorithm with, and with the possible number of target clusters to look for (we test for values ranging from~$2$ to~$10$). The related inertia plot is shown in Figure~\ref{fig:inertia-plot}, which clearly suggests $5$ as the optimal number of clusters to look for. We therefore save the KMeans++ clustering obtained for such~$5$ groups, and plot the t-SNE projection of our points in Figure~\ref{fig:tsne}, where the points are color-coded by the KMeans++ clustering labels. Interestingly, we can see that there is one central cluster to which the majority of points belong, while the other four clusters are likely to relate to outliers or unusually extreme combinations of MAV and related features (and indeed lie at the extremities of the plot). 

Figure \ref{fig:summary-stats} summarizes the features for each cluster, in order to help with their interpretation and discussion. For each group, we report the number of points belonging to it, but also show the average value of each feature that was considered during our clustering phase. In general, we can make the following observations:
\begin{itemize}
    \item Group 0: it comprises the great majority of data points, and points to average/ordinary evolution of misalignment conditions.
    \item Group 1: this cluster is characterised by much higher than average gas fees witnessed by the possible arbitrageurs. It is then sensible that we have e.g. low V0\_on\_vol\_usage, since the trading network is likely to be crowded.
    \item Group 2: extremely high MAV opportunities belong to this group, which are probably caused by exogenous further triggers (and indeed we see non-negligible time to decay).
    \item Group 3: it comprises MAV events that survive for extreme lengths of time, meaning that such arbitrage opportunities are not really
    exploited. This could be caused by moments of significantly low usage of the trading venue, which is also suggested by the very low gas fees witnessed around these events.
    \item Group 4: this cluster is actually made of only one data point. It shows the lowest time to decay, and an extreme amount of volume to be traded against the volume previously witnesses. Therefore, the possible arbitrageur would not be worried of adversarial trading.
\end{itemize}


\subsection{Regression Analysis}

We choose MAV events belonging to Group~$0$ as the data points of our interest for the incoming analysis. In this way, we try to focus on more ordinary conditions of price misalignment and re-alignment, and avoid exogenous extreme outliers.
We investigate the Ordinary Least Squares~(OLS) regression model, in which \texttt{clean\_MAV}, \texttt{avg\_gas}, and \texttt{Vmax\_on\_usage} are tested as explanatory features for \texttt{time\_decay}. Importantly, such features show negligible correlation among themselves.


After experimenting with feature selection and variable conversion, we find that \texttt{Vmax\_on\_usage} is not in general a statistically-significant feature. This could be caused by still overall low trading volume and liquidity on this specific venue, and we plan to test it again once we get further data (e.g. on layer-2 Uniswap v3 market venues). However, we still find the model proposed in Table~\ref{table-regression}, within which $x1=(clean\_MAV)^{-1/2}$ and $x2=(avg\_gas)^{-1/2}$. Such features have $p$-values that show significance at least at the~$0.15$ level, and the Durbin-Watson statistics is close to the desired magnitude of~$2$. Moreover, the model has a desirable $R^2$ of $0.19$, and indeed we see that it captures some variance in the distribution of decays as proposed with Figure~\ref{fig:true-pred}.

As expected, we find that MAV is negatively related to the decay time, while gas fees are positively related. Transformations of such variables (i.e. to the power of $-1/2$) are not intuitive, but are found necessary to obtain an initially significant relationship. Interestingly, gas fees have higher $t$-stats than MAV, which means that they have a higher influence on the decay time. We can interpret this by considering that MAV is an almost risk-free opportunity for profit, and so arbitrageurs could often be interested in collecting it despite its actual magnitude. Therefore, the estimate of gas fees to pay could in fact become a stronger determinant of the observed time to decay.
Finally, we remark that price differences can also decay due to exogenous reasons, i.e. people trading in the two contrary directions on the respective platforms without an explicit look for arbitrage.

\begin{figure}[tb]
\begin{center}
\setlength{\tabcolsep}{9pt}

\begin{tabular}{lclc}
\toprule
\textbf{Dep. Variable:}    &        y         & \textbf{  R-squared:         } &    0.190  \\
\textbf{Model:}            &       OLS        & \textbf{  Adj. R-squared:    } &    0.172  \\
\textbf{F-statistic:       } &    10.44   & \textbf{  Prob (F-statistic):} & 8.42e-05  \\
\bottomrule
\end{tabular}
\end{center}

\begin{center}
\setlength{\tabcolsep}{7pt}
\begin{tabular}{lcccccc}
\toprule
               & \textbf{coef} & \textbf{std err} & \textbf{t} & \textbf{P$> |$t$|$} & \textbf{[0.025} & \textbf{0.975]}  \\
\midrule
\textbf{x1}    &    -418.7701  &      277.568     &    -1.509  &         0.135        &     -970.291    &      132.751     \\
\textbf{x2}    &    1266.4030  &      277.568     &     4.563  &         0.000        &      714.882    &     1817.924     \\
\textbf{const} &   -3332.2671  &     1868.685     &    -1.783  &         0.078        &    -7045.304    &      380.770     \\
\bottomrule
\end{tabular}
\end{center}

\begin{center}
\setlength{\tabcolsep}{7pt}
\begin{tabular}{lclc}
\toprule
\textbf{Omnibus:}       & 38.134 & \textbf{  Durbin-Watson:     } &    1.854  \\
\textbf{Prob(Omnibus):} &  0.000 & \textbf{  Jarque-Bera (JB):  } &   72.444  \\
\textbf{Skew:}          &  1.682 & \textbf{  Prob(JB):          } & 1.86e-16  \\
\textbf{Kurtosis:}      &  5.753 & \textbf{  Cond. No.          } &     56.0  \\
\bottomrule
\end{tabular}
\caption{OLS model for the prediction of time of decay of price misalignments. Features have been standardised but not de-meaned.}
\label{table-regression}
\end{center}
\end{figure}

%% file: sections/09Conclusions.tex
\section{Related Work}
\label{sec:realted}
Substantial research in the area of arbitrage focuses on the identification of price discrepancies within DEXs. As an example,~\cite{berg2022empirical} studies market inefficiencies in Uniswap and Sushiswap and looks for cyclic arbitrage opportunities (i.e. optimal routing problem) and finds that markets struggled indeed to efficiently update prices during the initial trading volume explosion on DEXs in the late summer of~2020.
However, in our case we are interested in studying inefficiencies in DEXs' price formation especially against crypto CEXs.
The study in~\cite{barbon2023quality} compares the market quality of CEXs to DEXs, by discussing the microstructure of such exchanges. The authors analyse both transaction costs and deviations from the no-arbitrage condition, and find that CEXs provide indeed superior price efficiency but DEXs nicely eliminate custodian risk. Overall, they identify elevated gas prices and exchange fees as the main frictions harming DEX price efficiency, which further suggests to us to focus indeed on layer-2 blockchain AMMs for our study.
Then,~\cite{lehar2021decentralized} also carefully investigates the characteristics of trading and market design on Uniswap versus Binance. 
However, this study does not fully delve into price deviations among the exchanges, which is what we pursue in our study.

The Loss Versus Rebalancing (LVR) metric, as in \cite{milionis2023automated}, measures the losses incurred by liquidity providers (LPs) due to arbitrage activities rebalancing the AMM pools. LVR benchmarks the LP portfolio against a rebalancing portfolio that perfectly hedges the position's market risk. In contrast, the Maximal Arbitrage Value (MAV) defined in our paper quantifies the absolute profit of arbitrageurs, sometimes referred to as the net LVR per unit of measurement, such as per block or per minute. MAV diverges from LVR in two significant ways. Firstly, LVR operates under the assumption that price disparities between AMMs and CEXs are resolved within each block. While this may hold true for Uniswap pools on Ethereum, our observations indicate that on rollups like zkSync Era, the decay time of these price disparities extends over several minutes and spans multiple blocks. Our methodology accommodates this extended decay time, preventing the double-counting of the same opportunities. Secondly, LVR and rebalancing portfolios measure the losses of LPs, factoring in impermanent loss, whereas the MAV metric focuses solely on the absolute profit of arbitrageurs.

The empirical work by Fritsch et al.~\cite{fritsch2024measuring} measured LVR on Ethereum for Uniswap (v2) and (v3), operating under the LVR assumption that price differences decay within each block. We extend the empirical research with the adjusted framework for rollups to avoid the potential double-counting of arbitrage opportunities. Heimbach et al.~\cite{heimbach2024nonatomic} focus on estimating executed non-atomic arbitrage through MEV auctions on Ethereum. Our research differs by examining rollups where MEV auctions are not yet possible. Also, we evaluate unexploited arbitrage opportunities in contrast to the realized arbitrage in~\cite{heimbach2024nonatomic}. In conclusion, our research, alongside the benchmarks provided by the aforementioned works, offers new insights into the differences in arbitrage dynamics between Ethereum and its rollups.

\section{Discussion}
\label{sec:discussion}

Arbitrage signals inefficiencies within financial systems, and in the context of AMMs on rollups, these inefficiencies may arise from trader concerns about the new blockchains and limited trust in new DEXs. However, with the increase in trading activities, L2 AMMs should attract the attention of arbitrageurs. With faster block production (2s compared to 12s on Ethereum) and significantly lower gas fees, arbitrageurs can act swiftly and exploit smaller price disparities on rollups.

Arbitrage benefits traders by balancing prices across different trading platforms, but it can negatively impact LPs, causing impermanent loss. The LVR metric is designed to compare the earnings of arbitrageurs and LPs instead of directly measuring arbitrage. While arbitrageurs take advantage of price differences, LPs look for profitable AMM pools. These roles demand unique skills, and both are crucial to the AMM ecosystem.

\section{Conclusions}
\label{sec:conclusions}

Given the gradual transition of much of the trading activity from Ethereum to its rollups, this paper marks an initial investigation of AMMs in the rollup ecosystem, particularly focusing on zkSync Era. The distinct characteristics of the rollups compared to Ethereum, such as faster block production times and the absence of MEV, affect trading strategies, including arbitrage.

This paper introduces a novel metric, Maximal Arbitrage Value~(MAV) that measures the theoretical profit that an arbitrageur can earn by exploiting price divergences between AMMs and crypto CEXs, by taking into account the price impact of related trades.
We then propose an empirical framework to quantify arbitrage opportunities as witnessed on zkSync Era. We find that the cumulative MAV value in the period from July to September~2023 at the USDC-ETH SyncSwap pool (with ${\sim}6,000,000$ swaps and~\$43.73M trading volume) amounts to~\$104.96K. Then, we proceed to cluster and characterize such arbitrage opportunities by studying statistical relationships among their time to decay, MAV magnitude, and market fees that arbitrageurs must account for.

This research carries implications for the ongoing discussion about rollup design, particularly with regard to enabling the possibility of MEV in the rollup sequencers. In fact, the value of arbitrage opportunities~($\sim0.24\%$ of the trading volume) in the mentioned liquidity pool allows one to estimate the revenues from possible MEV auctions.
Future efforts will focus on applying the proposed framework to other AMMs, such as Uniswap~v3 forks on zkSync Era, and AMMs on optimistic rollups to provide a comprehensive understanding of arbitrage dynamics within both optimistic and zk rollup environments. 

%% file: sections/97AppendixTheory.tex
\section{Appendix} \label{app:theory}

\paragraph{Arbitrage between a CLMM and CEX.} The conservation function of a Uniswap v3 pool is an aggregate of all the individual LPs' conservation functions for the different ticks, each dependent on the exchange rate range that the LP wants to provide their liquidity for. 

Ticks are present at each price $p(i)=1.0001^i, i\in [1,2,...]$.
Suppose a LP supplies liquidity ($x, y$) only to users swapping within a specific range of exchange rates defined by two surrounding consecutive ticks, i.e. $[P_a/\alpha, P_a\alpha]$ with $\alpha>1$ and $P_a = \frac{x}{y}$ as the current spot price.
By definition, there exists some $i$ such that $[P_a/\alpha, P_a\alpha] \rightarrow [p(i), p(i+1)]$, and so we refer to ($x, y$) as ($x_i, y_i$) for precision and sake of future generalization. Within this environment, the shape of the trading function is identical to the case of liquidity provision of equivalent reserves
\begin{equation}
    x_i^{equiv} = \frac{x_i}{1-1/\sqrt{\alpha}}, \text{   and   } y_i^{equiv} = \frac{y_i}{1-1/\sqrt{\alpha}}
\end{equation}
under Uniswap v2.

We now allow the proportion of reserves to vary within the price range defined by the two ticks under consideration, and for simplicity, define $r_{i,x},r_{i,y} \geq 0$ as such varying quantities. 
Clearly, there is an upper bound to the change of reserves due to the implied movement to the next adjacent tick bounds. This translates into $0\leq r_{i,x} \leq x_i\cdot(\sqrt{\alpha}+1)$ and $0\leq r_{i,y} \leq y_i\cdot(\sqrt{\alpha}+1)^2$.
Thus, we can derive (similarly as for Uniswap v2) the percentage price impact when trading within two ticks $[p(i), p(i+1)]$ on Uniswap v3~\cite{Xu2023}. This is
\begin{equation}
    \rho_i(\Delta y) = \frac{\Delta x}{r_{i,x} + \frac{x_i}{\sqrt{\alpha}-1}},
\end{equation}
with the constraint that $\Delta y\leq r_{i,y}$, since otherwise we deplete one of the two reserves and move to the next tick.
Thus, our MAV formula becomes 
\begin{equation}
    MAV_i =
    V_{i, max} \cdot (P_a-P_c) - V_{max} \cdot P_a \cdot \rho_i( V_{i, max}),
\end{equation}
where $p(i) \leq P_c \leq p(i+1)$, and which needs to be solved enforcing the constraint $\Delta y\leq r_{i,y}$.

If the price misalignment spans multiple ticks, one then needs to iteratively compute $MAV_i$ and sum the related profits until realignment, i.e.
\begin{equation}
    MAV_{Uniswap\_v3} = \sum_{i:P_a\in [p(i), p(i+1)]}^{i:P_c\in [p(i), p(i+1)]} MAV_i.
\end{equation}

\begin{figure}{r}{5cm}
  \centering
  \includegraphics[width=5cm]{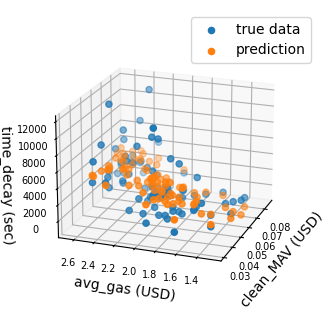} 
    \caption{Results of our regression model. The picture shows both predictions arising from our fitted regression, and the true data distribution to allow for the comparison of trends.}
    \label{fig:true-pred}
\end{figure}